\documentclass{article}
\usepackage{spconf,amsmath,graphicx}
\usepackage{amsmath,amssymb}
\usepackage{url}
\usepackage{xcolor}
\usepackage{multirow}

\usepackage{graphicx}
\usepackage{mathtools}
\usepackage{physics}
\usepackage{algpseudocode}
\usepackage{balance}
\usepackage{textcomp}
\usepackage{enumitem}
\usepackage{booktabs}
\usepackage{tabu}
\usepackage{float}
\usepackage{makecell}
\usepackage{hyperref}  
\hypersetup{
  colorlinks   = true, 
  urlcolor     = black, 
  linkcolor    = blue, 
  citecolor   = blue 
}

\usepackage[noadjust]{cite}
\usepackage{ctable} 

\makeatletter
\renewcommand\section{\@startsection {section}{1}{\z@}%
                                    {-1.8ex \@plus -0.1ex \@minus -.2ex}%
                                    {1.5ex \@plus1.12ex}%
                                    {\normalfont\Large\bfseries}}
\renewcommand\subsection{\@startsection{subsection}{2}{\z@}%
                                      {-2.0ex\@plus -0.0ex \@minus -.05ex}%
                                      {1.0ex \@plus 0.12ex}%
                                      {\normalfont\large\bfseries}}
\renewcommand\subsubsection{\@startsection{subsubsection}{3}{\z@}%
                                          {-1.2ex\@plus -0.5ex \@minus -.2ex}%
                                          {1.0ex \@plus .12ex}%
                                          {\normalfont\normalsize\bfseries}}

\usepackage{calrsfs}
\DeclareMathAlphabet{\pazocal}{OMS}{zplm}{m}{n}

\newcommand{\Lb}{\pazocal{L}}

\newcommand{\nCr}[2]{\,_{#1}C_{#2}} 
\usepackage{color}

\newcommand{\note}[1]{{\color{red}{\emph{#1}}}}
\newcommand{\point}[1]{{\color{blue}{\emph{#1}}}}
\newcommand{\tj}[1]{{\color{green}{\emph{#1}}}}
\renewcommand{\point}[1]{}
\renewcommand{\note}[1]{}
\renewcommand{\tj}[1]{}

\title{MULTI-SCALE SPEAKER DIARIZATION WITH NEURAL AFFINITY SCORE FUSION}
\name{Tae Jin Park, Manoj Kumar and Shrikanth Narayanan}
\address{University of Southern California}

\begin{document}
\ninept
%
\maketitle
\begin{abstract}
Identifying the identity of the speaker of short segments in human dialogue has been considered one of the most challenging problems in speech signal processing. Speaker representations of short speech segments tend to be unreliable, resulting in poor fidelity of speaker representations in tasks requiring speaker recognition. In this paper, we propose an unconventional method that tackles the trade-off between temporal resolution and the quality of the speaker representations. To find a set of weights that balance the scores from multiple temporal scales of segments, a neural affinity score fusion model is presented. Using the CALLHOME dataset, we show that our proposed multi-scale segmentation and integration approach can achieve a state-of-the-art diarization performance.

\end{abstract}
\section{Introduction}
\label{sec:intro}
Speaker diarization aims to cluster and label the regions in the temporal domain in terms of speaker identity. In general, the speaker diarization pipeline consists of speech activity detection  (SAD), segmentation, speaker representation extraction, and clustering. The segmentation process largely determines the accuracy of the final speaker label because the segmentation determines the unit of diarization output that cannot be altered during the clustering process. In terms of segmentation, a speaker representation faces an inevitable trade-off between the temporal accuracy and speaker representation quality. It has been shown in many previous studies that the speaker representation accuracy improves as the segment length increases \cite{snyder2017deep}. However, specifically in the context of speaker diarization, a longer segmentation means a lower resolution in the temporal domain because a segment is the unit of the process that determines the speaker identity. 

In the early days of speaker diarization, the clustering process was based on Bayesian information criterion (BIC) \cite{chen1998speaker}, which employs Mel-frequency cepstral coefficients (MFCCs) as a representation for speaker traits. With BIC-based clustering and MFCCs, speech segmentation techniques \cite{siegler1997automatic} with a variable segmentation length have been employed because the benefit of having a proper segment length for input speech outweighs the performance degradation from variable segment lengths. This trend has changed with the increase in newer speaker representation techniques, such as i-vector \cite{shum2013unsupervised, sell2014speaker} and x-vector \cite{snyder2018x, sell2018diarization}, where fixing the length of the segments improves the speaker representation quality and reduces additional variability. For this reason, many previous studies have made a point of compromise at 1.0 \cite{senoussaoui2013study} to 1.5 s \cite{sell2018diarization, landini2019but} depending on the domains they target. However, a fixed segment length has inherent limitations in terms of the temporal resolution because the clustering output can never be finer than the predetermined segment duration.

\begin{figure}[t]

\centerline{\includegraphics[width=8.1cm]{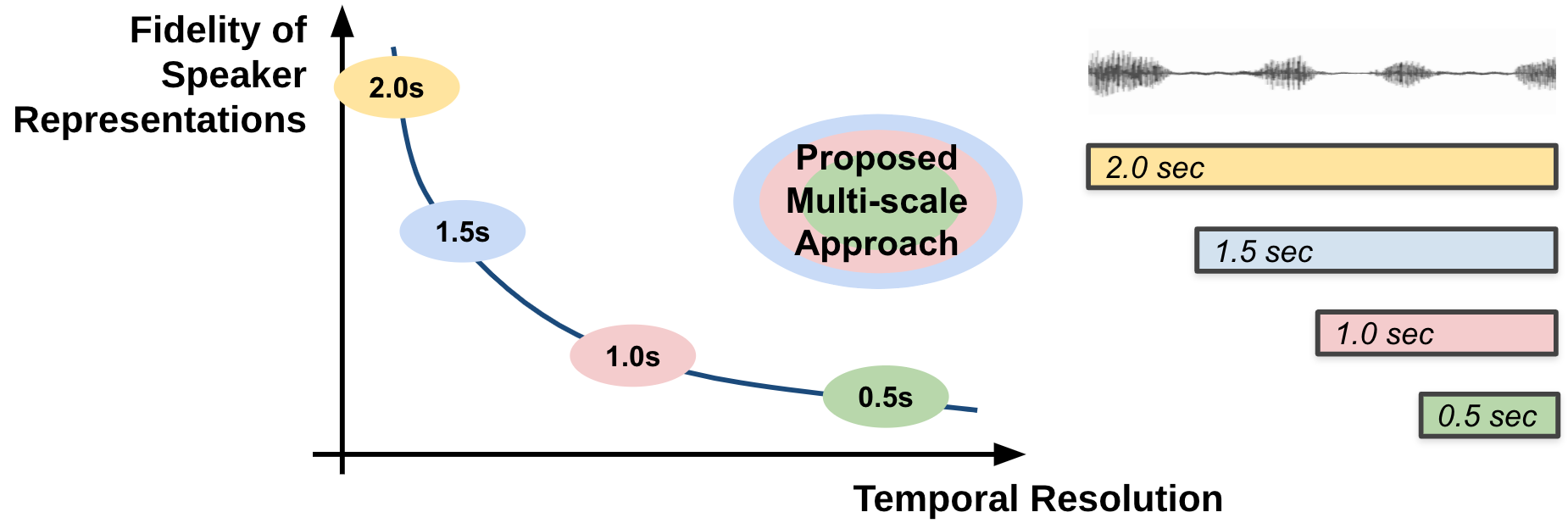}}
\vspace{-2.0ex}
\caption{Trade-off curve between fidelity of speaker representations and temporal resolution.}
\label{fig:trade_off}
\vspace{-5.0ex}
\end{figure}

Therefore, we propose a scheme that addresses the problem arising from such a trade-off and applies a new segmentation approach. The proposed method employs a multi-scale diarization solution where affinity scores from multiple scales of segmentation are fused using a neural score fusion system. The graph in Fig. \ref{fig:trade_off} shows the trade-off between segment length and fidelity of speaker representations from two segments. Our goal is for our system to be located on the graph above the trade-off curve with a higher temporal resolution while at the same time achieving a superior accuracy of the affinity measure.

There have been few studies related to the problem discussed herein. In terms of speaker embedding extraction, few studies have employed a multi-scale concept for speaker embedding \cite{jung2020multi, tang2019deep} targeting short utterance lengths. These studies apply multi-scale aggregation \cite{jung2020multi} or multilevel pooling \cite{tang2019deep} in the feature level in the neural network models. Because our proposed neural network model does not generate speaker embeddings, feature-level multi-scale approaches are far from our focus. 

By contrast, there are a few studies in which diarization systems aggregate the output of multiple modules. In \cite{huijbregts2009majority}, the authors employed a majority voting scheme on multiple segmentation streams. In \cite{bozonnet2010system}, the authors introduced a cluster matching procedure that can integrate multiple diarization systems. In addition, in \cite{stolcke2019dover}, a diarization output voting error reduction (DOVER) was presented for improving the diarization of a meeting speech. These previous studies deal with either a feature-level multi-scale concept of a neural network \cite{jung2020multi, tang2019deep} or a diarization system integration \cite{huijbregts2009majority, bozonnet2010system, stolcke2019dover}, whereas our proposed method focuses on the score fusion of multi-scale speech segments.

Our proposed approach has the following novelties. First, unlike conventional varying-length speech segmentation or single-scale segmentation modules, our system employs multiple discrete segment lengths and proposes a method to integrate the given scales. Second, the proposed method can attentively weigh the affinity from multiple scales depending on the domain and characteristics of the given speech signal. This distinguishes our work from approaches that require fusion parameters to be manually tuned on a development set. \cite{pardo2007speaker, yin2018analysis}. In addition to these novelties, our proposed multi-scale approach outperforms a single-scale diarization system and achieves a state-of-the-art performance on the CALLHOME diarization dataset. 

The remainder of this paper is structured as follows. In section 2, we introduce the segmentation scheme. In section 3, we introduce the proposed network architecture and how we train the proposed neural network. In section 4, we show the experimental results on various datasets and evaluation settings.

\section{Multi-Scale Diarization System}
\label{sec:format}
\subsection{Overview of the proposed system}
\begin{figure}[t]

\centerline{\includegraphics[width=8.1cm]{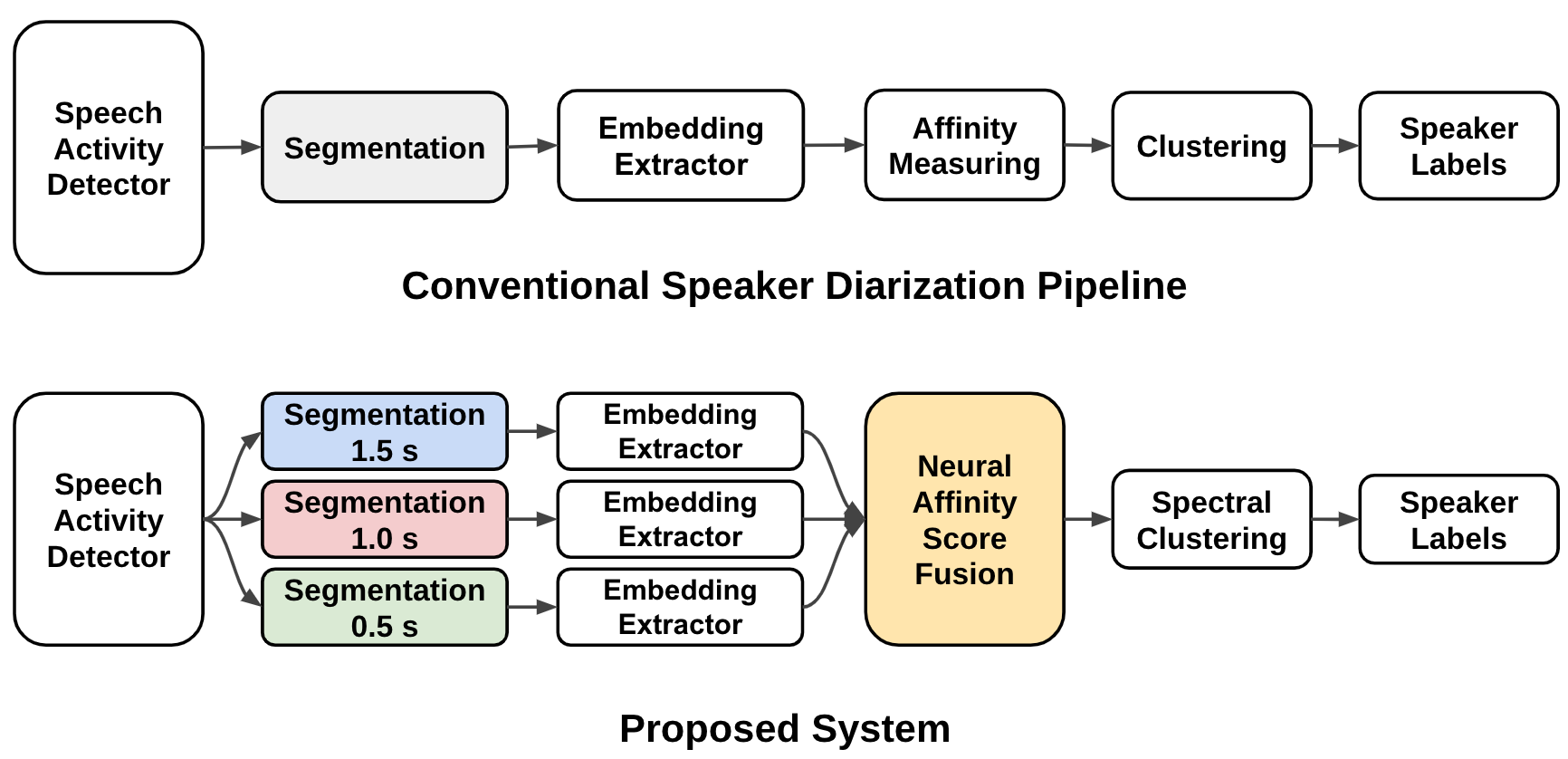}}
\vspace{-3.0ex}
\caption{Example of multi-scale segmentation scheme.}
\label{fig:data_flow}
\end{figure}

Fig. \ref{fig:data_flow} shows a block diagram of our proposed method as opposed to the conventional speaker diarization pipeline. For embedding the extractor, we employ an x-vector in \cite{snyder2018x, snyder_git}. We replace the segmentation process with a multi-scale segmentation process followed by a neural affinity score fusion (NASF) system, which will be described in the following sections. The NASF module outputs an affinity matrix similar to that in a conventional speaker diarization framework. In our proposed diarization, we employ the clustering method presented in \cite{park2019auto}. 
\subsection{Multi-scale segmentation}
\begin{figure}[t]
\centerline{\includegraphics[width=8.1cm]{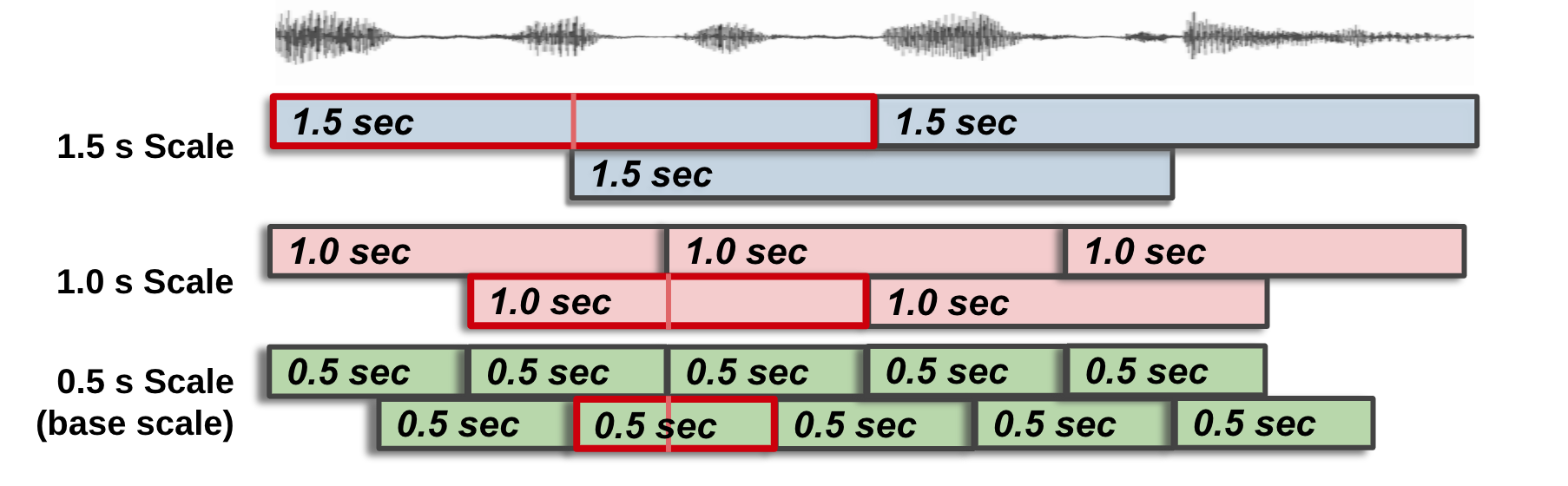}}
\vspace{-3.0ex}
\caption{Example of multi-scale segmentation and mapping scheme.}
\label{fig:overlap_segment}
\vspace{-2.0ex}
\end{figure}

Our proposed segmentation scheme for each scale is based on the segmentation scheme that appeared in a previous study \cite{sell2018diarization, snyder_git}. Fig. \ref{fig:overlap_segment} shows how our proposed multi-scale segmentation scheme works. Although many different scale lengths and numbers of scales can be adopted, we employ three different segment lengths: 1.5, 1.0, and 0.5 s. The hop-length is half the segment length, which is 0.75, 0.5, and 0.25 s, respectively. In addition, the minimum segment length of the each scale is set to 0.5, 0.25, and 0.17 s, respectively. 

We refer to the finest scale, 0.5 s, as the \textit{base scale} because the unit of clustering and labeling is determined by this scale. For each base scale segment, we select and group the segments from the lower temporal resolution scales (1.0 s and 1.5 s) whose centers are the closest to the center of the base scale segment. This mapping is shown by the red bounding boxes in Fig. \ref{fig:overlap_segment}. By selecting the segments as in Fig. \ref{fig:overlap_segment}, the clustering results are generated based on the base scale segments, whereas measuring the affinity for the clustering process is achieved using the distance obtained from all three scales.

\section{Neural Affinity Score Fusion Model}
\label{sec:neural_score_fusion_model}

For the speaker diarization task, learning an affinity fusion model is not a straightforward downstream task unlike training speaker embedding from speaker labels because the diarization output is obtained through a clustering (unsupervised learning) approach. Thus, we derived an indirect method that can learn a model for estimating the desirable weights for the affinity scores from multiple scales.

\subsection{Creating Data Labels}
\label{ssec:creating_data_labels}
\begin{figure}[t]
\centerline{\includegraphics[width=8.1cm]{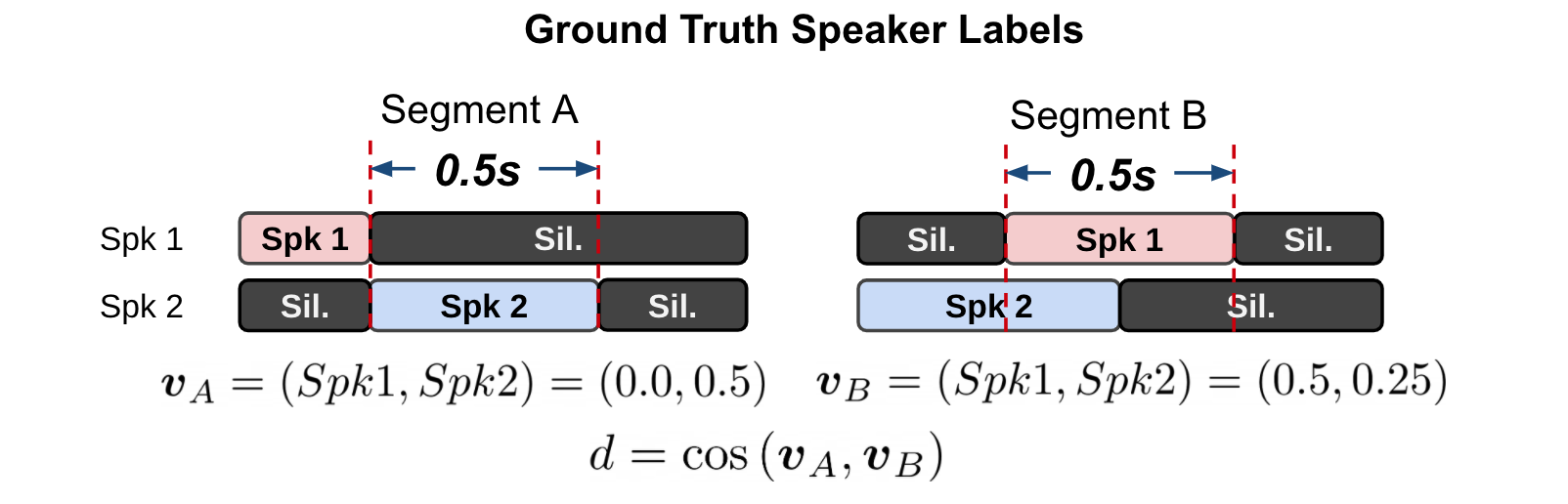}}
\vspace{-3.0ex}
\caption{Example of training data label generation.}
\label{fig:creating_labels}
\vspace{-3.0ex}
\end{figure}

To represent the ground-truth composition of the speakers in the given segments, we employ a concept of a \textit{speaker label vector} based on the duration of each speaker. The dimensions of the speaker label vector are determined based on the total number of speakers in a session. Fig.  \ref{fig:creating_labels} shows an example of how we create labels of training data. Let segments A and B be a pair of segments for which we want to obtain an affinity score label. In Fig. \ref{fig:creating_labels}, the speaker label vector $v_{A}$ obtains values of $(0, 0.5)$ and $(0.5, 0.25)$ from the duration of the speaker labels from segments A and B, respectively. 

Because the speaker label vectors are always positive, the ground truth cosine distance value ranges from zero to one. To match the range, the cosine similarity value from the speaker embeddings are min-max normalized to the (0, 1) scale. In total, for $L$ segments in the given session, we obtain $\nCr{L}{2}$ ground truth affinity score labels, which were created for the base scale, which has a segment length of 0.5 s.

\subsection{Affinity Score Fusion Networks}
\label{ssec:affinity_score_fusion_network}

To tackle the affinity weighting task, we employ a neural network model optimized using the mean square error (MSE) between the ground truth cosine similarity $d$ and weighted cosine similarity value $y$. We expect the estimated weight to minimize the gap between the ideal cosine similarity and the weighted sum of the given cosine similarity values $(c_1, c_2, c_3)$. To achieve this, we employ an architecture similar to that of a Siamese network \cite{koch2015siamese}, which shares the weights of the networks to process the two different streams of information. Thus, we build a neural network model that can capture the non-linear relationship between a set of affinity weights and a pair of speaker representations by setting up a pair of cloned neural networks.

\begin{figure}[t]
\centerline{\includegraphics[width=8.1cm]{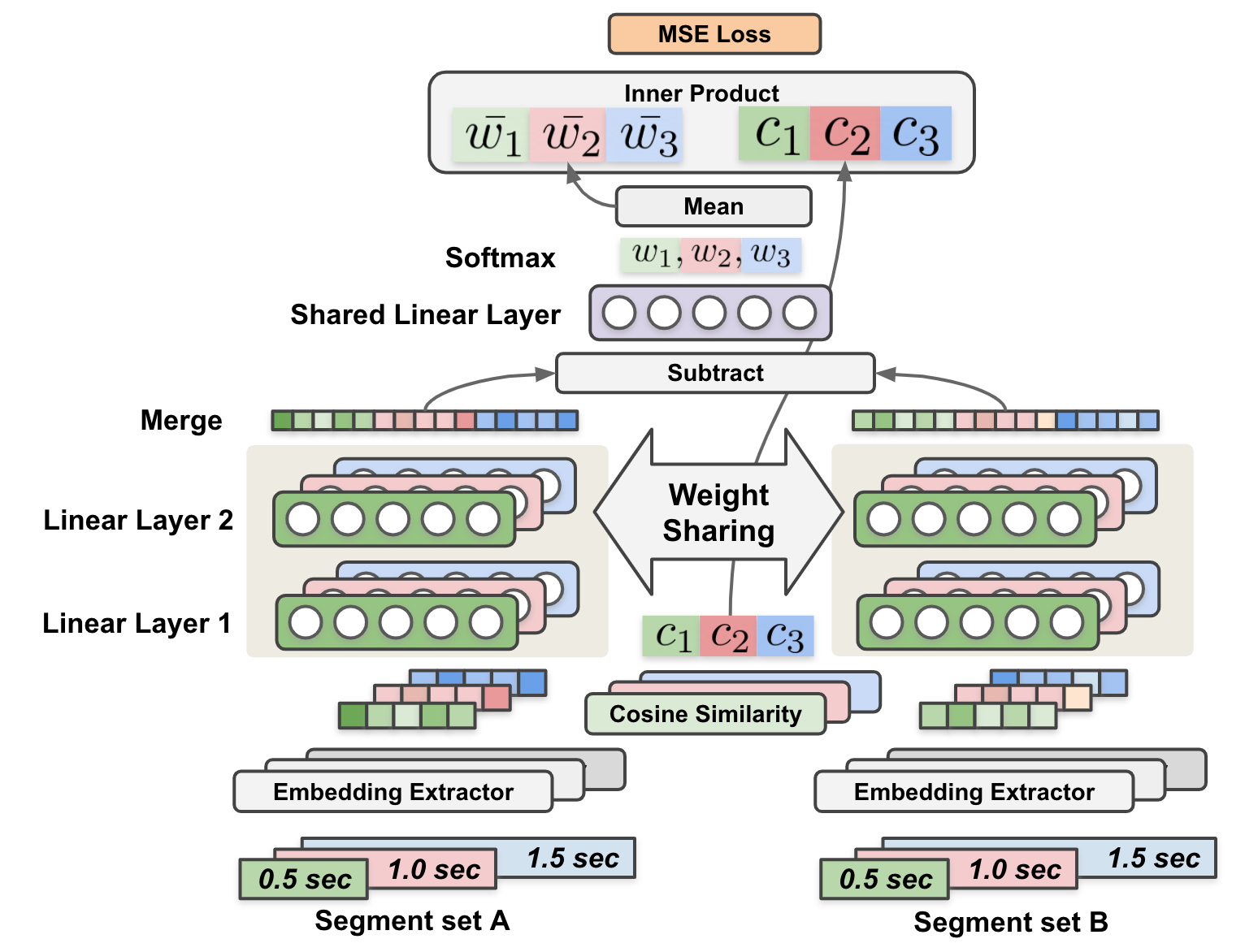}}
\vspace{-0.0ex}
\caption{Neural multi-scale score fusion model}
\label{fig:cos_weight_net}
\vspace{-3.0ex}
\end{figure}

Fig. \ref{fig:cos_weight_net} shows the architecture of the affinity score fusion network. After the multi-scale segmentation process, embeddings for each scale are extracted with three different embeddings for the three scales. The set of embeddings (segment set A) are then processed using three parallel multi-layer perceptrons (MLPs) and the output of the MLPs are merged to form an embedding from all three scales. The forward propagation of the input layer to the merging layer is also applied to another set of segments (segment set B) to obtain a merged embedding for this set. After forward propagation of two streams of input, the difference between two merged embeddings are passed to the shared linear layer, which outputs the softmax values. We then take the mean of the softmax values from $N$ input pairs.
\begin{equation}
    \textbf{w} = \Bigg(\frac{1}{N}\sum_{n=1}^{N} w_{1,n}, \frac{1}{N}\sum_{n=1}^{N} w_{2,n}, \frac{1}{N}\sum_{n=1}^{N} w_{3,n}\Bigg) 
\end{equation}
The set of averaged softmax values $\textbf{w}=(\bar{w}_1, \bar{w}_2, \bar{w}_3)$ weights the cosine similarity values, $\textbf{c}=(c_1, c_2, c_3)$, which are calculated using the speaker representations to obtain the weighted cosine similarity value as follows:
\begin{equation}
  y_{n} = \sum_{i=1}^{3} \bar{\omega}_{i} c_{i, n} = \textbf{w}^{T} \textbf{c}_{n}
\end{equation}
where $y_n$ is the output of the affinity weight network for the $n$-th pair out of $N$ pairs. Finally, the MSE loss is calculated using the ground truth cosine similarity value $d$ as follows:

\begin{equation}
  \Lb(\textbf{y}, \textbf{d}) = \frac{1}{N}\sum_{n=1}^{N} (y_{n} - d_{n})^{2}, 
\end{equation}
where $d_n$ is the $n$-th ground-truth cosine score for the $n$-th pair of segments. In inference mode, we also take the mean of $N$ sets of softmax values to obtain a weight vector $\textbf{w}$.

\subsection{Weighted Sum of Affinity Scores}
\label{ssec:weighted_sum_of_affinity_scores}

\begin{figure}[t]
\centerline{\includegraphics[width=8.1cm]{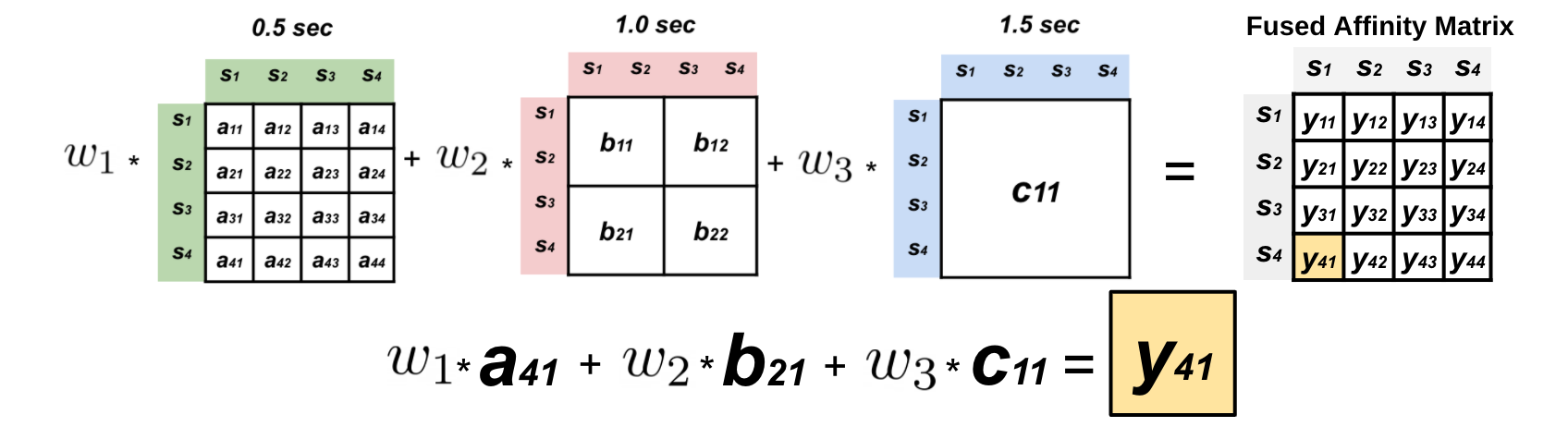}}
\vspace{-3.0ex}
\caption{Example of weighted sum of affinity matrices}
\label{fig:weighted_sum_mat}
\vspace{-3.0ex}
\end{figure}

Our proposed system estimates a weight vector $\textbf{w}$ for each input session (an independent audio clip under a real-world scenario). For inference of the affinity weight, we randomly select $N$=$5\cdot 10^{5}$ samples out of $\nCr{L}{2}$ pairs per session, which has $L$ base scale segments, and weigh the given affinity matrices as indicated in Fig. \ref{fig:weighted_sum_mat}. The weighted affinity matrix is then passed to the clustering module.

\subsection{Affinity Matrix and Clustering}
\label{ssec:affinity_matrix_and_clustering}

In our previous study \cite{park2019auto}, we showed that the cosine similarity when applying the NME-SC method can outperform that of the prevailing clustering approaches, such as a probabilistic linear discriminant analysis (PLDA) coupled with agglomerative hierarchical clustering (AHC). Thus, we employ cosine similarity and NME-SC method in \cite{park2019auto} to verify the efficacy of the proposed multi-scale affinity weight model by showing the additional improvement from the results in \cite{park2019auto}. In addition, we compare the performance with systems based on single-scale segmentation methods.

\section{EXPERIMENTAL RESULTS}
\label{sec:experimental results}

\subsection{Datasets}
\subsubsection{CALLHOME (NIST SRE 2000)} NIST SRE 2000 (LDC2001S97) is the most widely used diarization evaluation dataset and is referred to as CALLHOME. To compare its performance with performance of previous studies \cite{snyder2018x, snyder_git}, a 2-fold cross validation is conducted for evaluation on CALLHOME for AHC coupled with the PLDA method.
\subsubsection{Call Home American English Speech (CHAES)} CHAES (LDC97S42) is a corpus that contains only English speech data. CHAES is divided into train (80), dev (20), and eval (20) splits. 
\subsubsection{AMI meeting corpus} The AMI database consists of meeting recordings from multiple sites. We evaluated our proposed systems on the subset of the AMI corpus, which is a commonly used evaluation set that has appeared in numerous previous studies, and we followed the splits (train, dev, and eval) applied in these studies \cite{pal2020speaker, sun2019speaker, yella2015comparison}. 

\subsection{Training of NASF network}
For the training of our proposed neural network model, we use the CHAES- and AMI-train splits. We also apply the CHAES-dev and AMI-dev sets to tune the hyper-parameters of the network. We use MLPs with 2 hidden layers and 128 nodes each and apply the Adam optimizer with a learning rate of 0.001. 

\subsection{Distance measure and clustering algorithm}
\label{ssec:baselines}
In this paper, all systems employ a speaker embedding extractor (x-vector) that appeared in \cite{snyder2018x, snyder_git}. The following baselines are for the distance measure and clustering method.
\subsubsection{PLDA+AHC}
This approach is based on the AHC algorithm coupled with the PLDA as it appeared in \cite{snyder2018x, snyder_git}. The stopping criterion of the AHC was selected based on a grid-search for each development set. We use the PLDA model provided in \cite{snyder_git}.

\subsubsection{COS+NME-SC}
As stated in \cite{park2019auto}, NME-SC does not require a development set to tune the clustering algorithm. We use the same set of segments and speaker embeddings as PLDA+AHC but replace the distance measure with NASF from three different scales and replace the clustering algorithm with NME-SC. In this study, we do not evaluate combinations such as PLDA+NME-SC or COS+AHC because such combinations of algorithms have under-performed PLDA+AHC and COS+NME-SC \cite{park2019second}.

\begin{table*}[t]
\caption{Experimental results of baselines and the proposed methods}
\vspace{0.0ex}
\small
\label{tab:baselines}
\begin{center}
\begin{tabular}{ r |  c || c | c | c | c || c | c |  c || c | c | c } 
 \Xhline{3\arrayrulewidth}
& Number of & \multicolumn{3}{c|}{PLDA+AHC}   & Previous & \multicolumn{3}{c||}{COS+NME-SC}  & \multicolumn{3}{c}{Multi-scale COS+NME-SC}                                     \\
 Dataset & Sessions & $0.5 s$  &  $1.0 s$ &  $1.5 s$ & Studies  & $0.5 s$  &  $1.0 s$ &  $1.5 s$  & Equal Weight & NASF-D & NASF-S \\
\Xhline{3\arrayrulewidth}
AMI & 12  & 38.42 & 20.07 & 10.55 & 8.92 \cite{pal2020speaker} & 26.96 & 9.82 & 3.37  & 6.51 & 3.89 & \textbf{3.32}  \\
CHAES-eval & 20 & 4.58 & 3.15 & 3.28 & 2.48 \cite{park2019auto}& 8.71 & 3.35 & 2.48  & 2.52 & 2.47 & \textbf{2.04}  \\
CALLHOME & 500 & 17.89 & 9.13 & 8.39  & 6.63 \cite{lin2019lstm} & 20.96 & 7.81 & 7.29  & 6.64 & 7.02 & \textbf{6.46}  \\

\Xhline{3\arrayrulewidth}
\end{tabular}
\end{center}
\vspace{-5.0ex}
\end{table*}

\subsection{Diarization evaluation}

\subsubsection{Inference Setup}
\begin{itemize}[leftmargin=2.5ex]

    \item Equal Weight: This system is evaluated to show the efficacy of the NASF method over naive cosine similarity averaging. An equal weight system does not use any inference and applies equal affinity weights $(\frac{1}{3}, \frac{1}{3}, \frac{1}{3})$ for all sessions in all datasets.
    
    \item NASF-D: This system divides the input session into three equal-length sub-sessions and estimates six different affinity weight vectors ($\textbf{w}$) for the six different affinity matrices ($3! = 6$), which are intra-sub-session (three sessions) and inter-sub-session (three sessions) affinity matrices. Finally, we calculate the weighted sum of these matrices and join the affinity matrices to cluster the integrated matrix as a single affinity matrix.  

    \item NASF-S: This system estimates a set of affinity weights for an entire session. Thus, we have one affinity weight vector $\textbf{w}$ for each session, and the entire affinity matrix is weighted using this single weight vector.

\end{itemize}

\subsubsection{DER calculation}
\label{ssec:DER_calculation}
To gauge the performance of speaker diarization accuracy, we use oracle SAD output that excludes the effect of SAD module. For all evaluations and datasets, we estimate the number of speakers in the given session without additional information about speaker numbers. We employ an evaluation scheme and software that appeared in \cite{fiscus2006rich} to calculate  Diarization Error Rate (DER).

\begin{figure}[t]
\centerline{\includegraphics[width=7.6cm]{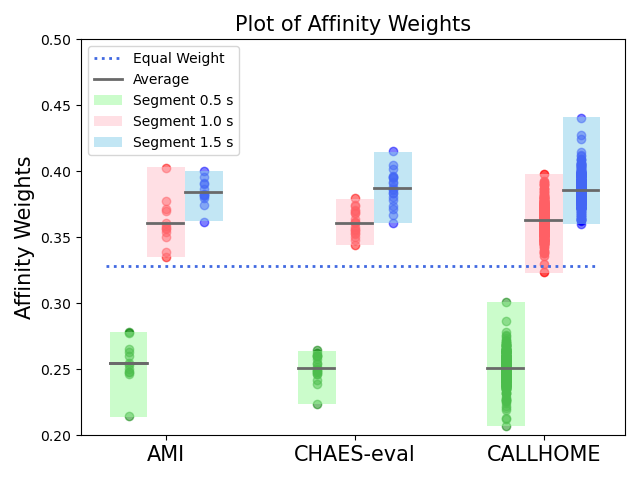}}
\vspace{-3.0ex}
\caption{Plot of affinity weights by datasets.}
\label{fig:weight_scatter_plot}
\vspace{-3.0ex}
\end{figure}

\subsection{Discussions}
We compare the DER obtained from the proposed method with the DER values obtained from each segment scale. Table \ref{tab:baselines} shows the DER from numerous settings and datasets. We show the DER values of the PLDA+AHC approach for three different segment scales (1.5, 1.0, and 0.5 s) and how the performance of the diarization changes with the distance measure and clustering method. We also list the lowest DER value that we could find that has appeared in a published paper on speaker diarization \cite{pal2020speaker, lin2019lstm}, including the CHAES-eval results of our previous study \cite{park2019auto}.

Most importantly, we compare the COS+NME-SC methods with segment lengths of 0.5, 1.0, and 1.5 s with the proposed method. The best performing system, NASF-S, obtains relative improvements with error rates of 1.5\%, 17.3\%, and 11.4\% for AMI, CHAES-eval, and CALLHOME, respectively, over the 1.5-s COS+NME-SC baseline. For the AMI corpus, the improvement was minor whereas the CALLHOME and CHAES-eval sets showed a significant improvement given that the DER result from COS-NME-SC with 1.5-s segments is already competitive compared to the results appearing in previous published studies.  
In Fig. \ref{fig:weight_scatter_plot}, we show the ranges and averages of the estimated weights over the sessions in each dataset. We can see that only the CALLHOME dataset shows a range that includes equal weight within the weight range for 1.0-s segments, whereas the weight ranges from AMI and CHAES-eval show no overlap with an equal weight. We conjecture that this is related to the result in which an equal weight shows an improvement for only CALLHOME.

From the experiment results, we can obtain a few valuable insights. The equal weight experiment gives conflicting results for AMI and CALLHOME. Nevertheless, from the equal weight experiment, we can verify that the desirable affinity weight cannot be simply found by averaging it and that the NASF approach can be a solution for estimating the desirable weights. 

The difference in performance gains between AMI and CALLHOME also shows the characteristics of a multi-scale approach. Because the longest segment we employ in our system is 1.5 s, we can argue that the DER reduction comes from the higher resolution of the segments. This becomes clear if we compare the proposed method with the DER we obtain from 0.5-s segments. However, the gain from our proposed method was not that significant with the AMI corpus. We speculate that this is caused by the characteristics of the dataset because the average length of the continuous speaker homogeneous region in the AMI corpus is 2.56 s, whereas the lengths for CALLHOME and CHAES are 2.17 and 2.07 s, respectively. In this sense, we can argue that the CALLHOME and CHAES datasets are more likely to benefit from the proposed multi-scale diarization approach because a higher resolution can capture shorter speaker homogeneous regions. 

Another important finding obtained from this study is that varying the affinity weights in a session (i.e., a diarization session that is being clustered) does not lead to a good performance. Having a constant affinity weight in a single affinity matrix leads to a better performance, as we can see from the NASF-S outperforming NASF-D. 

\section{Conclusions}
We proposed a method that mitigates the limit of the trade-off between the temporal resolution and the fidelity of the speaker representations. The proposed method estimates a set of weights that minimizes the gap between the weighted sum of the cosine affinity and the ground-truth affinity between a pair of segments. The proposed NASF system has a temporal resolution of 0.5 s and improves the diarization performance over conventional single-scale systems, achieving a state-of-the-art performance on the CALLHOME dataset. We believe that the proposed multi-scale score fusion approach on a diarization task can achieve a breakthrough in such research.

Further studies will include an online version of multi-scale speaker diarization where we can find the weight of the affinities in an online fashion. We expect our proposed multi-scale diarization framework to be applicable to many different diarization studies because our method is compatible with other modules in the diarization pipeline.

\vfill
\pagebreak

\bibliographystyle{IEEEbib}
\bibliography{strings,refs}

\begin{thebibliography}{10}

\bibitem{snyder2017deep}
David Snyder, Daniel Garcia-Romero, Daniel Povey, and Sanjeev Khudanpur,
\newblock ``Deep neural network embeddings for text-independent speaker
  verification.,''
\newblock in {\em Interspeech}, 2017, pp. 999--1003.

\bibitem{chen1998speaker}
Scott Chen, Ponani Gopalakrishnan, et~al.,
\newblock ``Speaker, environment and channel change detection and clustering
  via the bayesian information criterion,''
\newblock in {\em Proc. DARPA broadcast news transcription and understanding
  workshop}. Virginia, USA, 1998, vol.~8, pp. 127--132.

\bibitem{siegler1997automatic}
Matthew~A Siegler, Uday Jain, Bhiksha Raj, and Richard~M Stern,
\newblock ``Automatic segmentation, classification and clustering of broadcast
  news audio,''
\newblock in {\em Proc. DARPA speech recognition workshop}, 1997, vol. 1997.

\bibitem{shum2013unsupervised}
Stephen~H Shum, Najim Dehak, R{\'e}da Dehak, and James~R Glass,
\newblock ``Unsupervised methods for speaker diarization: An integrated and
  iterative approach,''
\newblock {\em IEEE Trans. Audio, Speech, Lang. Process.}, vol. 21, no. 10, pp.
  2015--2028, May 2013.

\bibitem{sell2014speaker}
Gregory Sell and Daniel Garcia-Romero,
\newblock ``Speaker diarization with plda i-vector scoring and unsupervised
  calibration,''
\newblock in {\em 2014 IEEE Spoken Language Technology Workshop (SLT)}. IEEE,
  2014, pp. 413--417.

\bibitem{snyder2018x}
David Snyder, Daniel Garcia-Romero, Gregory Sell, Daniel Povey, and Sanjeev
  Khudanpur,
\newblock ``X-vectors: Robust {DNN} embeddings for speaker recognition,''
\newblock in {\em Proc. IEEE Int. Conf. Acoust., Speech, Signal Process.}, Apr.
  2018, pp. 5329--5333.

\bibitem{sell2018diarization}
Gregory Sell, David Snyder, Alan McCree, Daniel Garcia-Romero, Jes{\'u}s
  Villalba, Matthew Maciejewski, Vimal Manohar, Najim Dehak, Daniel Povey,
  Shinji Watanabe, et~al.,
\newblock ``Diarization is hard: Some experiences and lessons learned for the
  {JHU} team in the inaugural {DIHARD} challenge.,''
\newblock in {\em Proc. INTERSPEECH}, Sep. 2018, pp. 2808--2812.

\bibitem{senoussaoui2013study}
Mohammed Senoussaoui, Patrick Kenny, Themos Stafylakis, and Pierre Dumouchel,
\newblock ``A study of the cosine distance-based mean shift for telephone
  speech diarization,''
\newblock {\em IEEE/ACM Transactions on Audio, Speech, and Language
  Processing}, vol. 22, no. 1, pp. 217--227, 2013.

\bibitem{landini2019but}
Federico Landini, Shuai Wang, Mireia Diez, Luk{\'a}{\v{s}} Burget, Pavel
  Mat{\v{e}}jka, Kate{\v{r}}ina {\v{Z}}mol{\'\i}kov{\'a}, Ladislav
  Mo{\v{s}}ner, Old{\v{r}}ich Plchot, Ond{\v{r}}ej Novotn{\`y}, Hossein
  Zeinali, et~al.,
\newblock ``But system description for dihard speech diarization challenge
  2019,''
\newblock {\em arXiv preprint arXiv:1910.08847}, 2019.

\bibitem{jung2020multi}
Youngmoon Jung, Seongmin Kye, Yeunju Choi, Myunghun Jung, and Hoirin Kim,
\newblock ``Multi-scale aggregation using feature pyramid module for
  text-independent speaker verification,''
\newblock {\em arXiv preprint arXiv:2004.03194}, 2020.

\bibitem{tang2019deep}
Yun Tang, Guohong Ding, Jing Huang, Xiaodong He, and Bowen Zhou,
\newblock ``Deep speaker embedding learning with multi-level pooling for
  text-independent speaker verification,''
\newblock in {\em ICASSP 2019-2019 IEEE International Conference on Acoustics,
  Speech and Signal Processing (ICASSP)}. IEEE, 2019, pp. 6116--6120.

\bibitem{huijbregts2009majority}
MAH Huijbregts, David~A van Leeuwen, and FM~Jong,
\newblock ``The majority wins: a method for combining speaker diarization
  systems,''
\newblock {\em Proc. INTERSPEECH}, 2009.

\bibitem{bozonnet2010system}
Simon Bozonnet, Nicholas Evans, Xavier Anguera, Oriol Vinyals, Gerald
  Friedland, and Corinne Fredouille,
\newblock ``System output combination for improved speaker diarization,''
\newblock in {\em Eleventh Annual Conference of the International Speech
  Communication Association}, 2010.

\bibitem{stolcke2019dover}
Andreas Stolcke and Takuya Yoshioka,
\newblock ``Dover: A method for combining diarization outputs,''
\newblock in {\em 2019 IEEE Automatic Speech Recognition and Understanding
  Workshop (ASRU)}. IEEE, 2019, pp. 757--763.

\bibitem{pardo2007speaker}
Jose Pardo, Xavier Anguera, and Chuck Wooters,
\newblock ``Speaker diarization for multiple-distant-microphone meetings using
  several sources of information,''
\newblock {\em IEEE Transactions on Computers}, vol. 56, no. 9, pp. 1212--1224,
  2007.

\bibitem{yin2018analysis}
Bing Yin, Jun Du, Lei Sun, Xueyang Zhang, Shan He, Zhenhua Ling, Guoping Hu,
  and Wu~Guo,
\newblock ``An analysis of speaker diarization fusion methods for the first
  dihard challenge,''
\newblock in {\em 2018 Asia-Pacific Signal and Information Processing
  Association Annual Summit and Conference (APSIPA ASC)}. IEEE, 2018, pp.
  1473--1477.

\bibitem{snyder_git}
David Snyder,
\newblock ``Callhome diarization recipe using x-vectors,'' Github, May 4, 2018.
  [Online]. Available:
  \url{https://david-ryan-snyder.github.io/2018/05/04/model_callhome_diarization_v2.html},
  [Accessed Oct. 21, 2020].

\bibitem{park2019auto}
Tae~Jin Park, Kyu~J Han, Manoj Kumar, and Shrikanth Narayanan,
\newblock ``Auto-tuning spectral clustering for speaker diarization using
  normalized maximum eigengap,''
\newblock {\em IEEE Signal Processing Letters}, vol. 27, pp. 381--385, 2019.

\bibitem{koch2015siamese}
Gregory Koch, Richard Zemel, and Ruslan Salakhutdinov,
\newblock ``Siamese neural networks for one-shot image recognition,''
\newblock in {\em Proceedings of the Workshop on Deep Learning in International
  Conference on Machine Learning, ICML}, 2015.

\bibitem{pal2020speaker}
Monisankha Pal, Manoj Kumar, Raghuveer Peri, Tae~Jin Park, So~Hyun Kim,
  Catherine Lord, Somer Bishop, and Shrikanth Narayanan,
\newblock ``Speaker diarization using latent space clustering in generative
  adversarial network,''
\newblock in {\em ICASSP 2020-2020 IEEE International Conference on Acoustics,
  Speech and Signal Processing (ICASSP)}. IEEE, 2020, pp. 6504--6508.

\bibitem{sun2019speaker}
Guangzhi Sun, Chao Zhang, and Philip~C Woodland,
\newblock ``Speaker diarisation using 2d self-attentive combination of
  embeddings,''
\newblock in {\em ICASSP 2019-2019 IEEE International Conference on Acoustics,
  Speech and Signal Processing (ICASSP)}. IEEE, 2019, pp. 5801--5805.

\bibitem{yella2015comparison}
Sree~Harsha Yella and Andreas Stolcke,
\newblock ``A comparison of neural network feature transforms for speaker
  diarization,''
\newblock in {\em Sixteenth Annual Conference of the International Speech
  Communication Association}, 2015.

\bibitem{park2019second}
Tae~Jin Park, Manoj Kumar, Nikolaos Flemotomos, Monisankha Pal, Raghuveer Peri,
  Rimita Lahiri, Panayiotis Georgiou, and Shrikanth Narayanan,
\newblock ``The second dihard challenge: System description for usc-sail
  team,''
\newblock {\em Proc. INTERSPEECH}, pp. 998--1002, 2019.

\bibitem{lin2019lstm}
Qingjian Lin, Ruiqing Yin, Ming Li, Herv{\'e} Bredin, and Claude Barras,
\newblock ``Lstm based similarity measurement with spectral clustering for
  speaker diarization,''
\newblock {\em Proc. INTERSPEECH}, 2019.

\bibitem{fiscus2006rich}
Jonathan~G Fiscus, Jerome Ajot, Martial Michel, and John~S Garofolo,
\newblock ``The rich transcription 2006 spring meeting recognition
  evaluation,''
\newblock in {\em Proc. Int. Workshop Mach. Learn. Multimodal Interaction}, May
  2006, pp. 309--322.

\end{thebibliography}
\end{document}